\documentclass{ws-jnopm}

\usepackage{graphicx}

\begin{document}
 
\markboth{C. -C. Lee et al.}{Ultra-Short Optical Pulse Generation with Single-Layer Graphene}

\catchline{}{}{}{}{}

\title{ULTRA-SHORT OPTICAL PULSE GENERATION WITH SINGLE-LAYER GRAPHENE}


\author{C. -C. LEE, and T. R. SCHIBLI}


\address{Department of Physics, University of Colorado\\Boulder, CO 80309, USA\\ChienChung.Lee@Colorado.EDU}


\author{G. ACOSTA, and J. S. BUNCH}

\address{Department of Mechanical Engineering, University of Colorado\\Boulder, CO 80309, USA}

\maketitle

\begin{history}
\end{history}

\begin{abstract}
Pulses as short as 260 fs have been generated in a diode-pumped low-gain Er:Yb:glass laser by exploiting the nonlinear optical response of single-layer graphene. The application of this novel material to solid-state bulk lasers opens up a way to compact and robust lasers with ultrahigh repetition rates.

\end{abstract}


\section{Introduction}	

Since the first report of single-layer graphene a few years ago, this material has already proven to possess a great number of interesting properties. One of these properties is its {non-linear} optical behavior, namely its {wavelength-independent} saturation of optical absorption. This saturable absorption can be utilized to induce {self-amplitude} modulation in a laser cavity, thus {mode-locking} lasers at wavelengths ranging from visible to far-infrared. In addition to the broadband absorption, graphene has an ultrafast relaxation time ($\sim$100 fs) in its carrier-carrier intraband scattering process and a slow relaxation time ($\sim$1 ps) in its carrier-phonon interband scattering process.\cite{pumpprobe} These two components make graphene excellent for generating self-starting, ultrashort optical pulses in solid-state lasers.

Thanks to the development of chemical vapor deposition ({CVD}) on copper foils,\cite{1} {large-area} {single-layer} graphene can be grown and transferred to substrates suitable for optics experiments. Multilayer graphene has been used to {mode-lock} fiber lasers,\cite{2,3} and very recently, to generate picosecond pulses in an Nd:YAG ceramics laser.\cite{4} {CVD-grown} {single-layer} graphene has also been demonstrated to be able to generate femtosecond pulses in a Cr:forsterite laser.\cite{5} While all these previous experiments used gain media with relatively high gain and high power pump sources, we built a low-gain Er:Yb:glass laser pumped with a compact diode laser, and we successfully showed that CVD-grown single-layer graphene has low enough insertion loss yet large enough saturable loss to achieve mode-locking. This work demonstrates that the low saturation intensity,\cite{2} low insertion loss, and high saturable loss of graphene can open up a way to compact, ultrahigh repetition rate lasers that have long been hoped for in robust frequency comb generation and other applications that require ultra stable pulsed light sources.

\section{Sample Preparation and Characterization}
\begin{figure}[th]
  \centering
    \includegraphics[width=\textwidth]{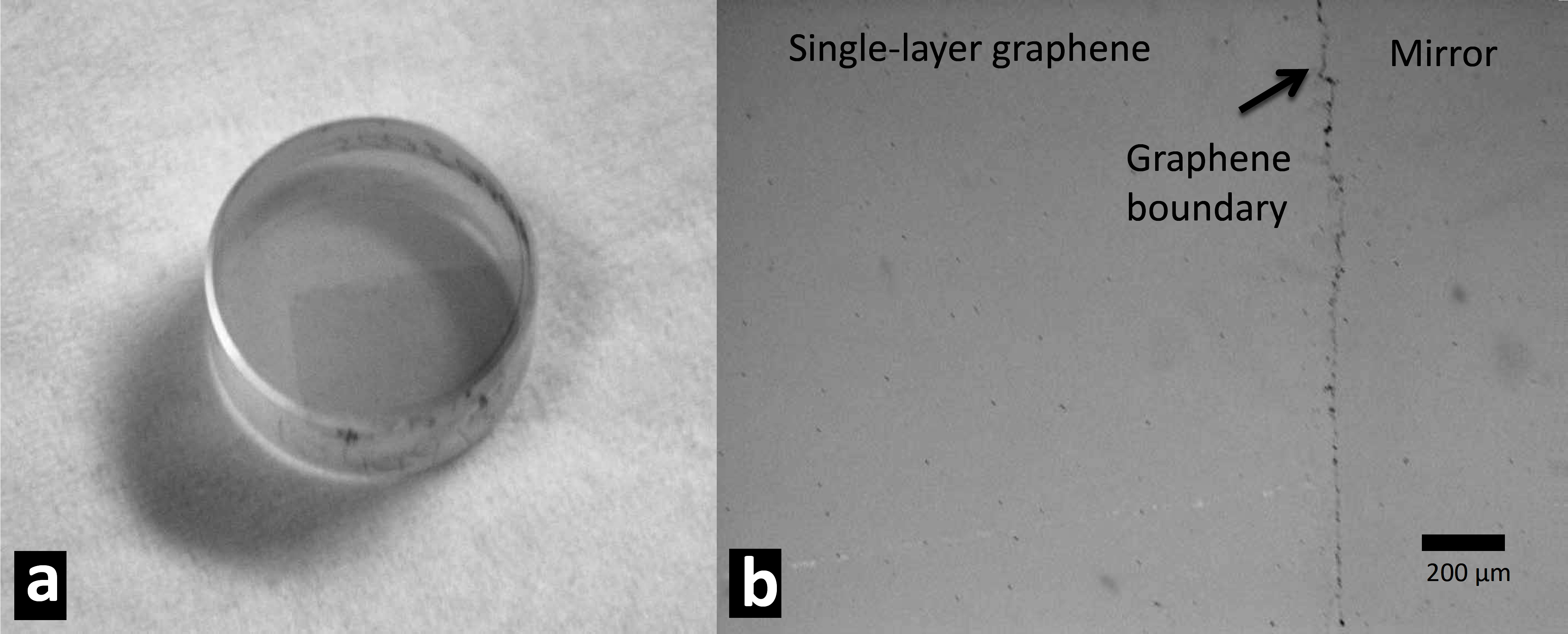}
  \caption{(a) Graphene saturable absorber mirror: Directly transferred graphene on a broadband laser mirror with $1/2$ inch diameter. The contrast of the picture is enhanced for clarity. (b) Optical microscopic view of single-layer graphene on a laser mirror showing good uniformity.}
  \label{grapheneSAM}
\end{figure}

The single-layer graphene, which was used as a saturable absorber in this experiment, was grown on the surface of copper foils by chemical vapor deposition. We choose to use copper over nickel due to its larger grain size and low solubility of carbon, which results in nearly 100\% of single-layer graphene.\cite{1} To have good surface quality and large grain size, the copper foils (Alfa Aesar \#13382, 25um thick) were first stripped of oxides in glacial acetic acid at 35$^\circ$C for five minutes,\cite{6} and then annealed at 1000$^\circ$C for 30 minutes in a quartz tube under a 500 sccm argon and a 65 sccm hydrogen flow. After that, the copper foils were maintained at 1000$^\circ$C and deposition of carbon atoms was achieved by flowing 500 sccm of argon, 65 sccm of hydrogen, and 50 sccm of methane at $\sim$830 mbar total pressure for five minutes. Then the tube was rapidly cooled down to room temperature by directly pulling it out from the tube furnace. In the cooling process, a flow of 500 sccm argon and 65 sccm hydrogen was maintained for about 90 minutes. The two sides of copper foils were then covered by single-layer graphene. In order to produce graphene on substrates that could be used for optical experiments, the copper foils were etched by an aqueous iron nitrate solution, and then the graphene sheets floating on the surface of the solution were transferred to DI water for cleaning. Finally, single-layer graphene sheets were directly transferred to a broadband laser mirror. Fig. \ref{grapheneSAM}(a) shows a photo of transferred graphene on a broadband, dielectric-coated mirror. Fig. \ref{grapheneSAM}(b) shows a microscopic picture of the transferred graphene, and it can be seen that graphene is uniform over most of the region, and only a few dark spots exist due to possibly multi-layer graphene or copper etchant residue. The uniformity of this large-area graphene reduces the non-saturable scattering loss of laser light. Also, focused laser light can be aligned on the sample easily without using a microscope, which is important for applications in solid-state bulk lasers.

The number of layers of graphene can be roughly identified by the contrast of image shown in a microscope. We further identified the number of layers by transmission electron microscopy and by transmission loss measurement. For the latter, a 980 nm laser beam with $\sim$60 $\mu$m diameter was scanned in one dimension along the sample while the transmission loss of graphene as a function of position was measured. The optical absorption of single-layer graphene can be shown by theory to be $\pi\alpha=2.3\%$, where $\alpha$ is the fine structure constant.\cite{7} Few-layer, independent graphene are then expected to give around an integer multiple of $2.3\%$ loss.

\begin{figure}[th]
  \centering
    \includegraphics[width=0.9\textwidth]{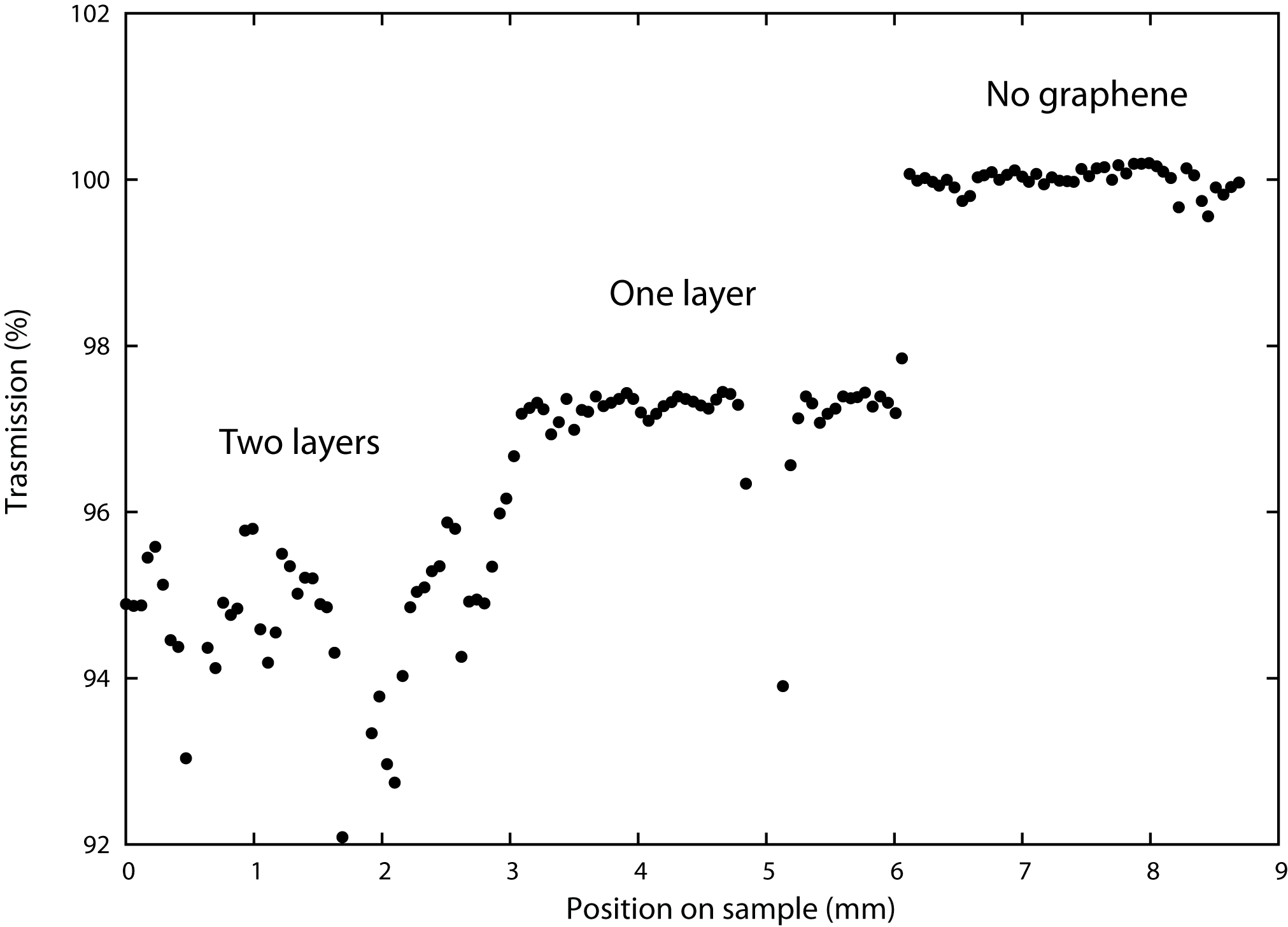}
  \caption{Transmission loss measurement of single and double-layer graphene on a broadband laser mirror. The single-layer and double-layer regions show
$2.7\pm0.1\%$ and $6.3\pm2.0\%$ transmission loss respectively. In the single-layer region, it can be seen there is a multilayer wrinkle of graphene, which has higher loss. In the double-layer region, the non-uniformity is caused by some contamination between the layers.}
  \label{trans}
\end{figure}

Fig. \ref{trans} shows a typical result of the transmission loss measurement. The graphene sample on a laser mirror has both single-layer region and double-layer region, which shows a transmission of $2.7\pm0.1\%$ and $6.3\pm2.0\%$, respectively. The extra loss in our sample can relate to scattering loss due to contamination and slight non-uniformity of the graphene. Despite the extra loss in the sample, this method is a quick way of examining the number of layers and knowing roughly the quality of graphene. The low loss provided by single-layer graphene is of great importance in low-gain solid-state lasers.

\section{Ultra-Short Pulse Generation}
\begin{figure}[th]
  \centering
    \includegraphics[width=1\textwidth]{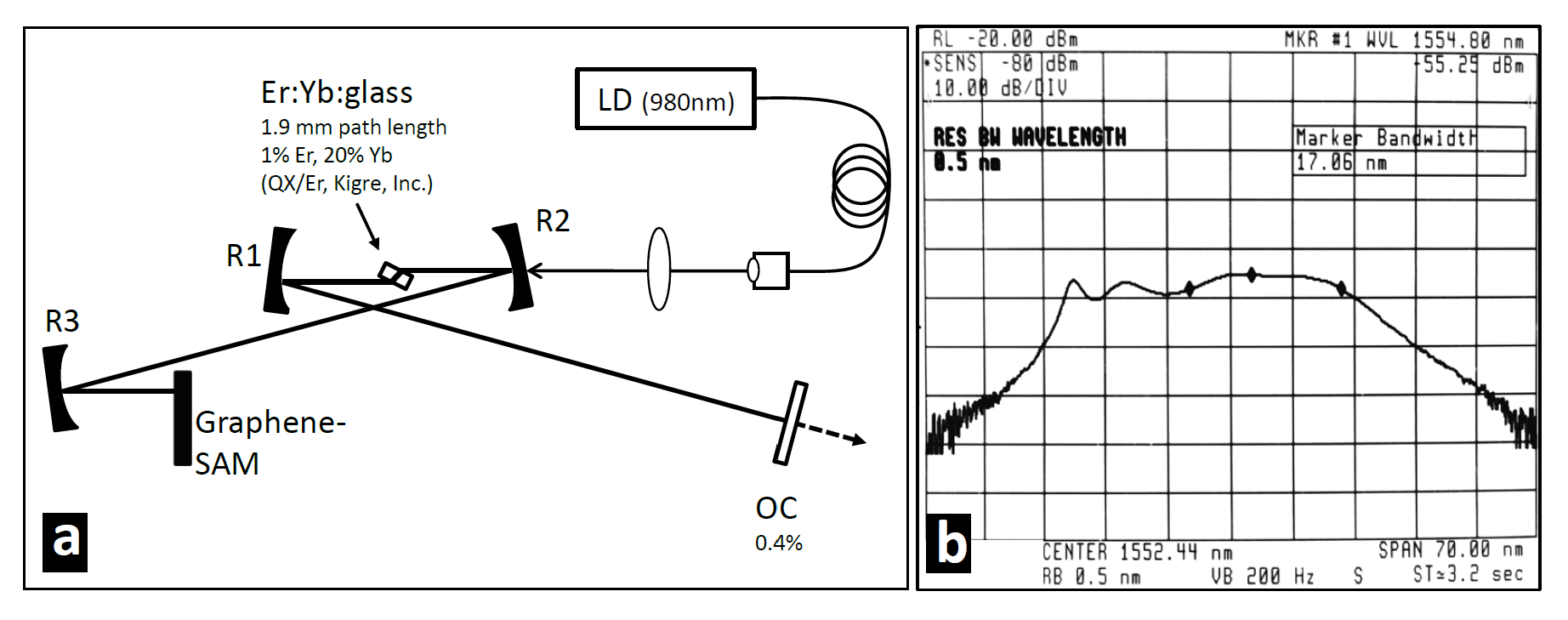}
  \caption{(a) Diode-pumped Er:Yb:glass laser cavity configuration. R1, R2, and R3 are concave broadband laser mirrors with ROC = 100 mm, 100 mm, and 50 mm, respectively. (b) Output optical spectrum. Spectral width $\sim$ 17 nm at a center wavelength of around 1550 nm. Resolution bandwidth: 0.5 nm.}
  \label{laser}
\end{figure}

To demonstrate the saturable absorption of our single-layer sample, we built a laser based on Er:Yb:glass (Kigre Inc.: QX/Er, 1\%Er, 20\%Yb, 1.6 mm plate under Brewster angle). The laser glass was directly diode pumped by a pigtailed single transversal mode and single wavelength 980 nm laser diode. The cavity was an astigmatically compensated, X-fold cavity with an additional focus on one of the end mirrors, which was replaced by the graphene-based saturable absorber mirror (Graphene-SAM), as shown in Fig. \ref{laser}(a). The chromatic dispersion of the cavity was dominated by the anomalous second-order dispersion of the laser glass. All cavity mirrors were commercial, low-dispersion broadband mirrors. The pulse repetition rate was 88 MHz. At an output coupling ratio of ~0.4\% and a pump power of about 130 mW stable mode-locking at 4.5 mW output power ($\sim$1.1W intracavity power) at a center wavelength of around 1550 nm was obtained. The mode-locked spectrum is shown in Fig. \ref{laser}(b). From an autocorrelation measurement, we inferred to a pulse duration of 260 fs, assuming a Gaussian pulse shape. The beam diameter on the saturable absorber was $\sim$30 $\mu$m, resulting in a maximum peak-intensity of $\sim$6 GW/cm$^2$.

The saturable loss of this graphene-based absorber was estimated to be about $\sim$20\% of the insertion loss based on the known cavity loss and the change of the laser average output power after mode-locking was initiated. Further characterization of the nonlinear absorption by balanced detection confirmed this number. Due to this limited modulation depth, the mode-locking process was not self-starting. The large non-saturable loss likely originates from lattice defects in the current graphene samples, and we believe that optimization of the CVD recipe and refinement of the sample preparation will lead to a much more favorable saturable-nonsaturable loss ratio.

\section{Conclusion}
We have shown that single-layer graphene grown by CVD on pure copper foils can be used as a saturable absorber. We have characterized our samples to show that we have single-layer graphene with high uniformity. Furthermore, we have generated short pulses in a low-gain solid-state laser operating at telecommunication wavelengths by mode-locking with the graphene saturable absorber mirror. The excellent ratio of saturable loss and low saturation intensity in single-layer graphene are extremely promising for for generating high-repetition rate pulse trains in low-noise solid-state lasers.

\section*{Acknowledgments}
We gratefully acknowledge the support from Dr. Kaoru Minoshima, AIST/NMIJ Tsukuba, Japan, who provided us the Er:Yb:glass sample that was used in this experiment.

\end{document}